# Quasi-one-dimensional ballistic ring in the field of circularly polarized electromagnetic wave

E.M. Epshtein[1], E.G. Fedorov[2], G.M. Shmelev[3]

[1] Institute of Radio Engineering and Electronics of the Russian Academy of Sciences, Fryazino, Russia; e-mail: eme253@ms.ire.rssi.ru

[2] Volgograd State University of Architecture and Civil Engineering, Volgograd, Russia; e-mail: eduard-f@mail.ru

[3] Volgograd State Pedagogical University, Volgograd, Russia; e-mail: shmelev@fizmat.vspu.ru

Dynamics is studied of an electron in a quasi-one-dimensional ballistic ring under circularly polarized electromagnetic field propagating along the normal to the ring plane. The average emission intensity from the ring is calculated. The value and direction of the electron average angular velocity in the ring depend on the incident wave parameters. It is found that the ring average dipole moment can remain constant under certain conditions. Possibility is shown of higher harmonics enhancement in the ring radiation spectrum.

Studying properties of quasi-one-dimensional rings is one of the most promising directions in physics of low-dimensional structures (e.g., [1, 2]). Incidentally, quantum phenomena were considered preferably. Meanwhile, quasi-one-dimensional rings have interesting *classical* electrodynamic properties, too [3 – 9]. In present work, we investigate electromagnetic response of a quasi-one-dimensional ring in a circularly polarized (CP) electromagnetic field propagating along the normal to the ring plane.

Consider a plane ring with width small compared to its radius $R$. Such a ring is a quantum well between two concentric potential barriers with quantum confinement along the radius. We assume that the electron mean free path is large in comparison to $2\pi R$, so that the electron executes ballistic classic motion along the ring circumference. Before the CP wave turning on at $t = 0$, the electron with energy $W$ moves along the ring circumference with constant angular velocity. Let a CP wave is incident on the ring with the wavelength large compared to the ring diameter. In dipole approximation, only the electric field of the wave

$$\boldsymbol{E} = E_0 \{\cos(\omega t + \beta), \sin(\omega t + \beta)\} \tag{1}$$

acts on the electrons.

The collisionless motion of the electrons interacting with the CP wave is described by equation

$$\ddot{\varphi} + \Omega^2 \sin(\varphi - \omega t - \alpha) = 0, \tag{2}$$

where $\varphi = \varphi(t)$ is angular coordinate counted off the $Ox$ axis, $\alpha = \beta \pm \pi$,

$$\Omega^2 = \frac{|e|E_0}{mR}, \tag{3}$$

$e$ and $m$ are electron charge and effective mass, respectively.

The solution of Eq. (2) is

$$\varphi(t) = \omega t + \alpha + 2 \begin{cases} \arcsin[k\,\text{sn}(\Psi_1, k)], & k < 1, \\ \arcsin[\tanh(\Psi_2)], & k = 1, \\ \arcsin[\text{sn}(\Psi_3, k^{-1})] & k > 1, \end{cases} \tag{4}$$

where

$$\Psi_1 = \xi \Omega t + F\left[\arcsin\left(k^{-1} \sin \frac{\varphi(0) - \alpha}{2}\right) k\right] \quad (k < 1), \tag{5}$$

$$\Psi_2 = \xi \Omega t + F\left(\frac{\varphi(0) - \alpha}{2}, 1\right), \tag{6}$$

$$\Psi_3 = k\xi \Omega t + F\left(\frac{\varphi(0) - \alpha}{2}, k^{-1}\right) \qquad (k > 1), \tag{7}$$

$$\xi = \text{sgn}(\dot{\varphi}(0) - \omega), \tag{8}$$

$$k = \left[\frac{[\dot{\varphi}(0) - \omega]^2}{4\Omega^2} + \frac{1}{2}[1 - \cos(\varphi(0) - \alpha)]\right]^{\frac{1}{2}}, \tag{9}$$

$\boldsymbol{j}(0) = \pm R^{-1}(2W/m)^{1/2}$ is angular velocity of the electron with energy $W$ at $t = 0$, $F(x, q)$ is incomplete elliptic integral of the first kind, $\text{sn}(x, q)$ is elliptic sine.

Let the initial conditions $\varphi(0)$, $\dot{\varphi}(0)$ be the same for all the $N$ electrons in the ring. The dipole moment with respect to the ring center and the radiation intensity are

$$\boldsymbol{P} = NeR\{\cos\varphi, \sin\varphi\} \tag{10}$$

and

$$I(t) = \frac{2}{3c^3}\left(\ddot{\boldsymbol{P}}\right)^2 = \frac{2}{3c^3}(NeR)^2\left(\dot{\varphi}^4 + \ddot{\varphi}^2\right), \tag{11}$$

respectively [10].

For an electron in the ring under CP wave acting, we have

$$\dot{\varphi}^4 = \omega^4 + 8\xi\omega^3\Omega G + 24\omega^2\Omega^2 G^2 + 32\xi\omega\Omega^3 G^3 + 16\Omega^4 G^4, \tag{12}$$

where

$$G = \begin{cases} k\,\mathrm{cn}(\Psi_1, k), & k < 1, \\ \mathrm{sech}(\Psi_2, k^{-1}), & k = 1, \\ k\,\mathrm{dn}(\Psi_3, k^{-1}), & k > 1, \end{cases} \tag{13}$$

$$\ddot{\varphi}^2 = 4\Omega^4 \begin{cases} k^2\,\mathrm{sn}^2(\Psi_1, k)\mathrm{dn}^2(\Psi_1, k), & k < 1, \\ \sinh^2(\Psi_2)\mathrm{sech}^4(\Psi_2), & k = 1, \\ \mathrm{sn}^2(\Psi_3, k^{-1})\mathrm{cn}^2(\Psi_3, k^{-1}), & k > 1, \end{cases} \tag{14}$$

$\mathrm{cn}(x, q)$ is elliptic cosine, $\mathrm{dn}(x, q)$ is delta amplitude.

The radiation fundamental frequency defined by Eq. (11) is equal to

$$\omega_0 = \Omega\pi \begin{cases} \dfrac{1}{2\boldsymbol{K}(k)}, & k < 1, \\ 0, & k = 1, \\ \dfrac{k}{\boldsymbol{K}(k^{-1})}, & k > 1, \end{cases} \tag{15}$$

where $\boldsymbol{K}(q)$ is complete elliptic integral of the first kind. The dependence of $\omega_0/\Omega$ on $q$ is shown in Fig. 1.

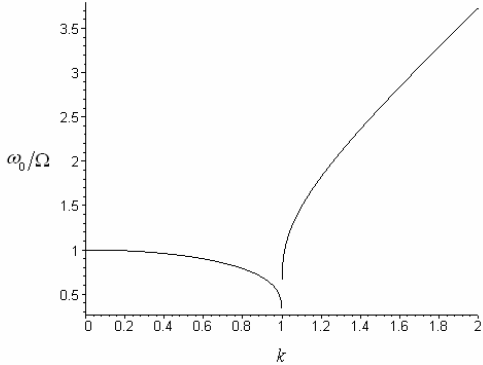

Fig. 1. The dependence of $\omega_0/\Omega$ on $q$.

It follows from Eq. (11) that the average radiation intensity from the ring is

$$\langle I \rangle \cong \frac{2}{3c^3}(NeR)^2 \begin{cases} \omega^4 + 2k^2\Omega^2\left(6\omega^2 + \Omega^2\right) + 6\Omega^4 k^4, & k \ll 1, \\ \omega^4, & k = 1, \\ \omega^4 + 8\xi\omega^3\Omega k + 24\omega^2\Omega^2 k^2 + 32\xi\omega\Omega^3 k^3 + 16\Omega^4 k^4, & k \gg 1, \end{cases} \quad (16)$$

where the angular brackets mean averaging over period $T = 2\pi/\omega_0$, $k$ being defined by Eq. (9).

At arbitrary $k$, the average intensity $\langle I \rangle$ can be found numerically from Eq. (11),

$$\langle I \rangle = T^{-1}\int_0^T I(t)dt \,. \quad (17)$$

If the following conditions fulfil

$$\begin{cases} \omega = \dot\varphi(0), \\ \beta = \varphi(0), \varphi(0) \pm \pi, \end{cases} \quad (18)$$

then $\boldsymbol{E}$ field is collinear to $\boldsymbol{P}$ vector at any instant, so that the electron moves along the ring circumference with constant angular velocity $\boldsymbol{j} = \boldsymbol{j}(0) = \boldsymbol{w}$. At that case, the radiation intensity is [10]

$$I = \frac{2}{3c^3}(NeR)^2\,\omega^4\,. \quad (19)$$

Note that the radiation intensity is $I_0 = (2/3)c^{-3}(NeR)^2\boldsymbol{j}^4(0)$ up to $t = 0$ instant.

According with Eq. (4), the electron angular velocity averaged over the period $T = 2\pi/\omega_0$ is

$$\langle \dot\varphi \rangle = \begin{cases} \omega, & k \le 1, \\ \omega + \xi\pi\Omega\dfrac{k}{K(k^{-1})}, & k > 1. \end{cases} \quad (20)$$

Under

$$\left| \langle \dot\varphi \rangle \right| \ll \frac{2\pi}{t^*} \quad (21)$$

condition, the electron displacement along the ring circumference is negligible in comparison to $2\pi R$ at $t^* \cong \nu \gg 2\pi\omega^{-1}, 2\pi\omega_0^{-1}$ ($\nu$ is electron collision frequency). During $t^*$ time, the average dipole moment $\langle \boldsymbol{P} \rangle$ remains almost constant under CP wave acting. The latter fact is confirmed with numerical solution of the equation of motion (2).

Let us make numerical estimates. At $R = 10^{-5}$ cm, $m = 0.1m_e$, $W = 10^{-3}$ eV, we have $\dot{\varphi}(0) = \pm 5.9 \cdot 10^{11} \text{c}^{-1}$. The dependence of $\langle I \rangle / I_0$ ratio on $k = k(\omega)$ found numerically at $E_0 = 3 \cdot 10^3$ V/cm, $\beta = 5\pi/4$, $\varphi(0) = \pi/4$ is shown in Fig. 2.

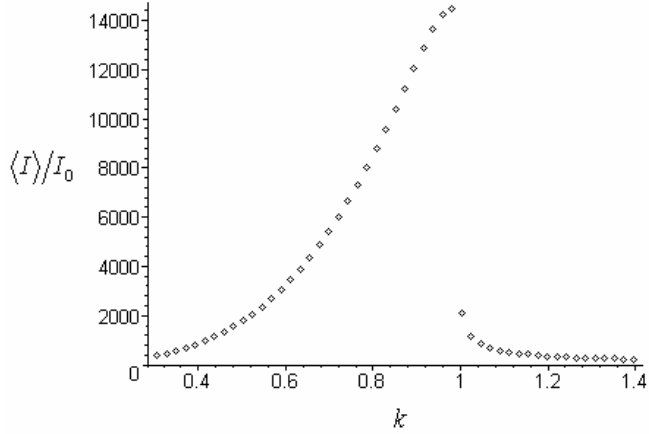

Fig. 2. $\langle I \rangle / I_0$ as a function of $k = k(\boldsymbol{w})$ with $E_0 = \text{const}$.

The dependence $\langle I \rangle / I_0$ on $k = k(E_0)$ at $\boldsymbol{w} = 1 \cdot 10^{12} \text{c}^{-1}$ and the same values of $\varphi(0)$, $\dot{\varphi}(0)$ and $\beta$ is shown in Fig. 3

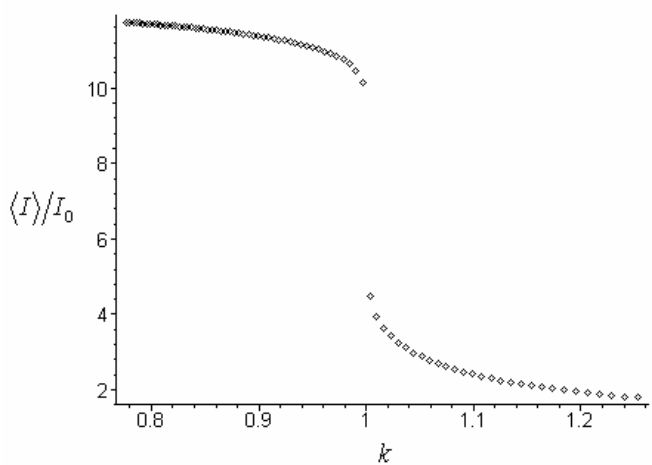

Fig.3. $\langle I \rangle / I_0$ as a function of $k = k(E_0)$ with $\boldsymbol{w} = \text{const}$.

The radiation spectrum at parameters indicated is shown in Fig. 4.

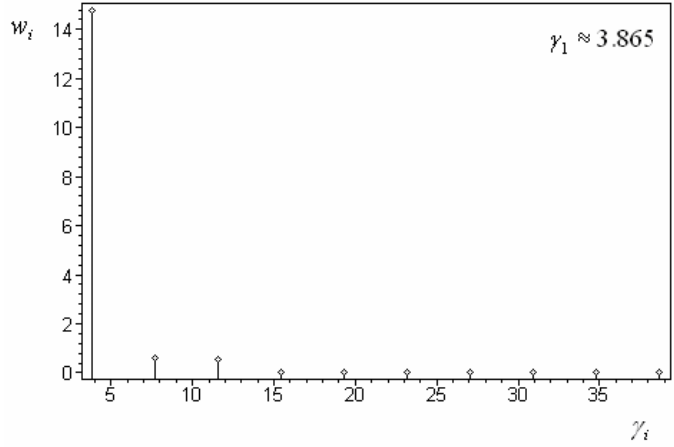

Fig. 4. The ring radiation spectrum ($\gamma_i = i\,\omega_0 / \big|\dot\varphi(0)\big|$).

Since $I(t)$ function is periodical at given parameters ($k \approx 0.089 \neq 1$), the spectrum is presented by a correspondence

$$i\omega_0 \rightarrow \sqrt{a_i^2 + b_i^2} \equiv w_i\,, \qquad i = 1,\,2,\,...,\qquad(22)$$

where $a_i$ and $b_i$ are coefficients in the $I(t)$ function expansion into Fourier series

$$I(t) = \langle I\rangle + I_0 \sum_{i=1}^{\infty}\big[a_i \cos(i\omega_0 t) + b_i \sin(i\omega_0 t)\big].\qquad(23)$$

The condition (21) fulfils, for example, at $E_0 = 3 \cdot 10^2$ V/cm, $\omega = 9.4\,10^{11}$ c$^{-1}$, $\beta = 5\,\pi/4$, $\boldsymbol{j}(0) = \boldsymbol{p}/4$, $\dot\varphi(0) = -5.9 \cdot 10^{11}$c$^{-1}$. In that case, $k \approx 1.06$, $\langle\dot\varphi\rangle \approx 3.4 \cdot 10^7$c$^{-1}$, $2\pi/t^* \approx 6.3 \cdot 10^{10}$ c$^{-1}$, $\omega_0 \approx 9.4\,10^{11}$ c$^{-1}$, $\langle\boldsymbol{P}\rangle/(NeR) = \big\{\langle\cos\varphi\rangle,\,\langle\sin\varphi\rangle\big\} \approx \{0.655,\,0.657\}$. The dependence of $\varphi$ on $\tau = t\big|\dot\varphi(0)\big|$ is shown in Fig. 5.

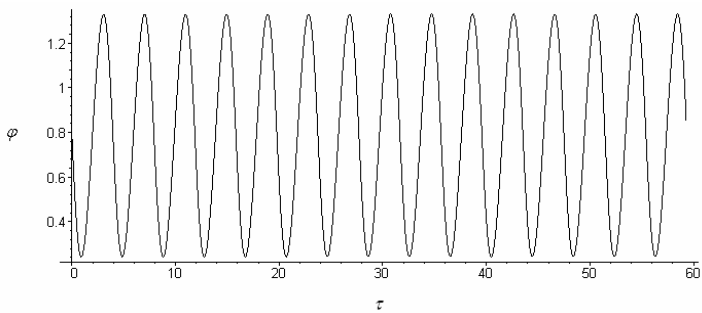

Fig. 5. The dependence of $\varphi$ on $\tau = t\big|\dot\varphi(0)\big|$.

The radiation spectrum in that situation is shown in Fig. 6.

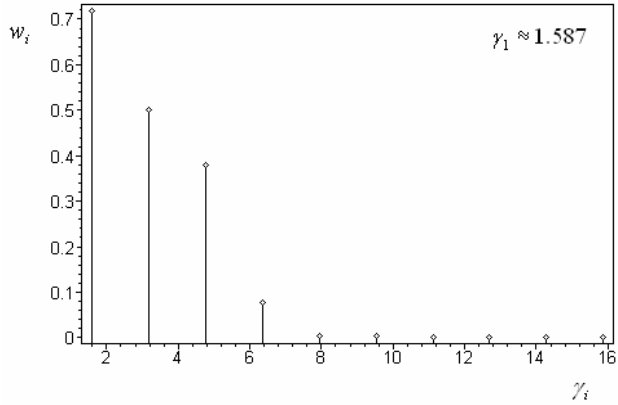

Fig. 6. The ring radiation spectrum under (21) condition. Enhancement is seen of higher harmonics in comparison to Fig. 4.

It is seen that the higher harmonics intensities decrease much more slowly in comparison with the situation corresponding to Fig. 4. Therefore, enhancement of higher harmonics occurs under condition (21) fulfillment.

At above-mentioned parameter values and collision frequency $\nu = 10^{10}\,\mathrm{c}^{-1}$ we have $\left|\dot{\varphi}(0)\right| \gg 2\pi\nu$, as well as $\omega$, $\omega_0 \gg 2\pi\nu$ at $k \neq 1$ and $\omega \gg 2\pi\nu$ at $k = 1$. It justifies the collisionless approximation used in description of the electron dynamics in the ring.

The work was supported partly by the Russian Fund of Basic Research (Project No. 02-02-16238).